\documentclass[12pt]{article}
\title{\bf Straight--line string in the einbein field formalism
}
\author{Yu.S.Kalashnikova\thanks{e-mail: yulia@vxitep.itep.ru},
A.V.Nefediev\thanks{e-mail: nefediev@vxitep.itep.ru}}
\date{\it Institute of Theoretical and Experimental Physics, 117259, Moscow,
Russia}
\newcommand{\ds}{\displaystyle}
\newcommand{\be}{\begin{equation}}
\newcommand{\ee}{\end{equation}}

\newcommand{\pcd}[1]{\mathstrut\partial #1}
\newcommand{\f}[1]{\varphi_{#1}}

\newcommand{\dbp}[2]{\{#1#2\}'}
\begin{document}
\maketitle

\begin{abstract}
The canonical description of the straight--line string is given in the einbein
field formalism. The system is quantized and Regge spectrum is reproduced.
The covariant analogue of the Newton--Wigner coordinate is found, and 
peculiarities of the gauge fixing in $\tau$-reparametrization group are 
discussed.
\end{abstract}

The idea that QCD at large distances is a string theory gives rise to a
relatively simple model of hadrons, which are represented by some 
configuration of the string with quarks or antiquarks at the ends. In the 
models of such a type ordinary mesons and baryons are described by the string
in its ground state, while excited string levels correspond to exotic 
$q\bar q g$, $q\bar q gg$ $\ldots$ mesons or $qqqg$, $qqqgg$ $\ldots$ 
baryons \cite{1}. Physically such a picture is rather appealing, but no general
solution for the problem of string with massive ends is known even at the 
classical level, because the radial and orbital quark motion cannot be 
separated from the pure string modes, that makes the problem very 
complicated. It is tempting, however, to use an ansatz to describe the ground 
state of the string, assuming the straigt--line form \cite{2} for the string 
part of the action: 
\be
S=S_{\rm quarks}+S_{\rm string},
\label{1}
\ee
\be
S_{\rm string}=
-\sigma\int_{\tau_1}^{\tau_2}d\tau
\int_0^1d\beta\sqrt{(\dot{w}w')^2-\dot{w}^2w'^2},
\label{2}
\ee
\be
w_{\mu}=(1-\beta)x_{1\mu}+\beta x_{2\mu},
\label{3}
\ee
where $x_{1\mu}(\tau)=w_{\mu}(\tau,\;0)$ and 
$x_{2\mu}(\tau)=w_{\mu}(\tau,\;1)$ are the coordinates of the endpoints, and
$\dot{w}_{\mu}=\frac{\partial w_{\mu}}{\partial\tau}$, 
$w'_{\mu}=\frac{\partial w_{\mu}}{\partial\beta}$.

Ansatz (\ref{3}) does not, in general, satisfy the string Euler--Lagrange 
equations with boundary conditions which correspond to placing quarks at
the string ends, as it was shown in detail in \cite{3}. Indeed, for the 
Euler--Lagrange equations to be respected, the world surface
$w_{\mu}(\tau,\beta)$ in (\ref{2}) should be the minimal one, while in 
accordance with the Catalan theorem \cite{4} the ruled surface (\ref{3})
is minimal only if it is either a plane or a helicoid. The latter posibility
suggests that for the case of large orbital momenta and lowest radial
excitations (leading trajectory), 
{\it i.e.} when the quark term in (\ref{1}) can be neglected, the theory given
by equations (\ref{2}) and (\ref{3}) is a good first approximation and
deserves some attention. 

In what follows we present the two-body treatment of theory (\ref{2}), (\ref{3}) 
in the framework of the einbein field formalism \cite{5} which allows to
separate the centre-of-mass motion and provides the natural environment
for the identification of the physical degrees of freedom.

Einbein fields were introduced to get rid of square roots which enter the
Lagrangians of relativistic systems, though at the price of introducing extra
dynamical variables. For example \cite{6}, the Lagrangian of a pointlike
particle,
\be
L=-m\sqrt{\dot{x}^2},
\label{4}
\ee
can be rewritten as
\be
L=-\frac{\mu\dot{x}^2}{2}-\frac{m^2}{2\mu},
\label{5}
\ee
where $\mu=\mu(\tau)$ is the einbein field, and the original form (\ref{4})
is recovered if the solution of the Euler--Lagrange equation for the 
einbein field $\mu$ is substituted into Lagrangian (\ref{5}). Form 
(\ref{5}) is quadratic in velocity, that provides an opportunity to
express it explicitly in terms of canonical momentum 
$p_{\mu}=\frac{\partial L}{\partial \dot{x}_{\mu}}$, and it is very helpful 
in the Hamiltonian formulation of the theory.

The extension of the method to the straight--line string is to introduce
a continious set of einbein fields $\nu(\beta)$, $0\le\beta\le 1$, rewriting
the Lagrangian from (\ref{2}) as 
\be
L=-\frac12\int_0^1d\beta\nu(\beta)\left(\dot{w}^2-\frac{(\dot{w}x)^2}{x^2}
\right)+\frac12\int_0^1d\beta\frac{\sigma^2x^2}{\nu(\beta)},
\label{6}
\ee
$$
x_{\mu}=x_{1\mu}-x_{2\mu}.
$$

The Euler--Lagrange equation for the field $\nu(\beta)$,
\be
\frac{\partial}{\partial\tau}\frac{\delta L}
{\delta\dot{\nu}(\beta)}-
\frac{\delta L}{\delta\nu(\beta)}=0,
\label{7}
\ee
reduces to the extremum condition 
\be
\frac{\delta L}{\delta\nu(\beta)}=0
\label{8}
\ee 
and has the solution
\be
\nu_{\rm extr}=-\frac{\sigma x}{\sqrt{\frac{(\dot{w}x)^2}{x^2}-\dot{w}^2}},
\ee
which returns us back to action (\ref{2}).

To separate the centre-of-mass motion we introduce a new set of variables
instead of $x_{1\mu}$ and $x_{2\mu}$:
\be
x_{\mu}=x_{1\mu}-x_{2\mu},\quad X_{\mu}=\zeta x_{1\mu}+(1-\zeta)x_{2\mu}
\ee
$$
\zeta=\frac{\int_0^1d\beta\nu\beta}{\int_0^1d\beta\nu}.
$$

In terms of these new variables Lagrangian (\ref{6}) takes the form
\be
L=-\frac12M\left(\dot{X}^2-\frac{(x\dot{X})^2}{x^2}\right)
-\frac12m\left(\dot{x}^2-\frac{(x\dot{x})^2}{x^2}\right)+\frac12kx^2
\ee
with the notations
\be
M=\int_0^1d\beta\nu,\quad m=\int_0^1d\beta\nu(\beta-\zeta)^2,\quad 
k=\int_0^1\frac{\sigma^2}{\nu}.
\label{11}
\ee

The canonical momenta are
$$
P_{\mu}=\frac{\ds \pcd{L}}{\ds \pcd{\dot{X_{\mu}}}}=-M\left(
\dot{X}_{\mu}-\frac{x_{\mu}(x\dot{X})}{x^2}\right),
$$
\be
p_{\mu}=\frac{\ds \pcd{L}}{\ds \pcd{\dot{x_{\mu}}}}=-m\left(
\dot{x}_{\mu}-\frac{x_{\mu}(x\dot{x})}{x^2}\right),
\label{12}
\ee
$$
\kappa(\beta)=\frac{\delta L}{\delta \dot{\nu}(\beta)}=0,
$$
and the Hamiltonian $H_0=(P\dot{X})+(p\dot{x})-L$ takes the form
\be
H_0=-\frac{P^2}{2M}-\frac{p^2}{2m}-\frac{kx^2}{2}.
\label{13}
\ee

At first glance the centre-of-mass motion is already separated out in 
Hamiltonian (\ref{13}), but it is not the case: first, the coefficients in 
(\ref{13}) depend on the einbein field via relations (\ref{11}), and,
second, primary constraints are present in the theory as it is easily seen
from expressions (\ref{12}) for the canonical momenta:
\be
\f{1}=(Px),\quad \f{2}=(px),\quad \f{3}(\beta)=\kappa(\beta).
\label{14}
\ee

So we deal with a constrained theory and should act along the lines of the 
general Dirac's procedure \cite{7}. First we are to define the full
Hamiltonian
\be
H=H_0+\Lambda \f{1}+\lambda \f{2}+\int_0^1d\beta e(\beta)\f{3}(\beta),
\ee
where constrains (\ref{14}) are added with Lagrange multipliers $\Lambda$,
$\lambda$ and $e(\beta)$. As we treat the einbeins as dynamical variables
the Poisson bracket is given by the expression
$$
\{AB\}=
\frac{\pcd{A}}{\pcd{P_{\mu}}}\frac{\pcd{B}}{\pcd{X_{\mu}}}-
\frac{\pcd{A}}{\pcd{X_{\mu}}}\frac{\pcd{B}}{\pcd{P_{\mu}}}+
\frac{\pcd{A}}{\pcd{p_{\mu}}}\frac{\pcd{B}}{\pcd{x_{\mu}}}-
\frac{\pcd{A}}{\pcd{x_{\mu}}}\frac{\pcd{B}}{\pcd{p_{\mu}}}+
\hspace*{2cm}
$$
\be
\hspace*{4cm}
+\int_0^1d\beta\left(\frac{\delta{A}}{\delta{\kappa(\beta)}}
\frac{\delta{B}}{\delta{\nu(\beta)}}-
\frac{\delta{A}}{\delta{\nu(\beta)}}
\frac{\delta{B}}{\delta{\kappa(\beta)}}\right).
\label{16}
\ee

Primary constraints (\ref{14}) give rise to the secondary constraints
$$
\f{4}=\{\f{1}H\}=\frac{(pP)}{m}-\lambda(Px),
$$
\be
\f{5}=\{\f{2}H\}=\frac{p^2}{m}-kx^2,
\label{17}
\ee
$$
\f{6}=\{\f{3}(\beta)H\}=\frac{P^2}{2M^2}+\frac{p^2}{2m^2}(\beta-\zeta)^2
+\frac{\sigma^2x^2}{2\nu^2(\beta)},
$$
and no further constraints appear, because equations $\{\f{a}H\}=0$, $a=4,5,6$,
define the Lagrange multipliers. At the constraints surface these equations are
\footnote{As usually sign $\approx$ denotes 
the so called \lq\lq weak" equality, {\it i.e.} equality which holds
when all constraints are set equal to zero.}
$$
\{\f{4}H\}\approx\Lambda\frac{P^2}{m}=0
$$
\be
\{\f{5}H\}\approx 2\lambda\frac{p^2}{m}+2\lambda kx^2+\int_0^1d\beta e(\beta)
\left(\frac{p^2}{m^2}(\beta-\zeta)^2-\frac{\sigma^2x^2}{\nu^2(\beta)}\right)=0
\label{18}
\ee
$$
\{\f{6}H\}\approx\lambda\frac{p^2}{m^2}(\beta-\zeta)-\lambda\frac{\sigma^2x^2}
{\nu^2(\beta)}+e(\beta)\frac{\sigma^2x^2}{\nu^3(\beta)}+\hspace*{3cm}
$$
$$
\hspace*{3cm}+\int_0^1d\beta' e(\beta')\left(\frac{P^2}{M^3}+\frac{p^2}{m^3}
(\beta-\zeta)^2(\beta'-\zeta)^2\right)=0.
$$

Solution of equations (\ref{18}) is
\be
\Lambda=0,\quad \lambda=0,\quad e(\beta)=e_0\nu(\beta),
\ee
where $e_0$ is an arbitrary coefficient. 
It is not surprising that the Lagrange
multipliers are defined only up to an arbitrary constant; initial action
(\ref{2}), (\ref{3}) is invariant under $\tau$-reparametrization 
transformations (and introducing the eibein does not spoil this invariance),
so we deal with a gauge theory. From general considerations \cite{7} it
means that there exist two linear combinations, one of primary constraints
(\ref{14}) and another of secodary ones (\ref{17}), which form a conjugated
pair of the first class constraints. This pair is easily identified to be
\be
\Phi_1=\int_0^1d\beta\nu(\beta)\f{3}(\beta),\quad
\Phi_2=\int_0^1d\beta\nu(\beta)\f{6}(\beta)\approx -H_0.
\label{20}
\ee 

Explicit calculations with bracket (\ref{16}) demonstrate that indeed
$\{\Phi_{1,2}\f{a}\}\approx 0$, $a=1,\ldots,6$.

Due to presence of the first class constraints (\ref{20}) the 
constraints matrix $C_{ab}=\{\f{a}\f{b}\}$ is degenerate $({\rm det}C=0)$, 
and this pair should be eliminated in calculation of Dirac brackets. 
Technically it is convenient to define the preliminary brackets first,
\be
\{AB\}'=\{AB\}-\sum_{i,j}\{A\f{i}\}C^{-1}_{ij}\{\f{j}B\},
\label{21}
\ee
where $i,j=1,2,4,5$ only. In what follows the physical variables are
constructed in terms of $P_{\mu}$, $X_{\mu}$, $p_{\mu}$ and $x_{\mu}$, which
have the following preliminary brackets:
\be
\begin{array}{l}
\dbp{P_{\mu}}{X_{\nu}}=g_{\mu\nu},\quad\dbp{P_{\mu}}{P_{\nu}}=
\dbp{P_{\mu}}{p_{\nu}}=\dbp{P_{\mu}}{x_{\nu}}=0\\
{}\\
\dbp{X_{\mu}}{X_{\nu}}=\frac{\ds S_{\mu\nu}}{\ds P^2}\\
{}\\
\dbp{X_{\mu}}{x_{\nu}}=\frac{\ds x_{\mu}P_{\nu}}{\ds P^2}\\
{}\\
\dbp{X_{\mu}}{p_{\nu}}=\frac{\ds p_{\mu}P_{\nu}}{\ds P^2}\\
{}\\
\dbp{p_{\mu}}{x_{\nu}}=g_{\mu\nu}-\frac{\ds 
P_{\mu}P_{\nu}}{\ds P^2}-\frac{\ds 2}{\ds mc}P_{\mu}p_{\nu}-
\frac{\ds 2k}{\ds c}x_{\mu}x_{\nu}\\
{}\\
\dbp{p_{\mu}}{p_{\nu}}=\frac{\ds 2k}{\ds c}S_{\mu\nu}\\
{}\\
\dbp{x_{\mu}}{x_{\nu}}=\frac{\ds 2}{\ds mc}S_{\mu\nu}
\end{array}
\label{22}
\ee
$$
S_{\mu\nu}=x_{\mu}p_{\nu}-x_{\nu}p_{\mu},\quad c=2kx^2+\frac{2p^2}{m}.
$$ 

The would-be physical variables are defined by means of the tetrade formalism 
\cite{8}. The tetrade of vectors is given by
\be
e_{0\mu}=\frac{P_{\mu}}{\sqrt{P^2}},\quad 
e_{i\mu}e_{j\mu}=-\delta_{ij},
\quad e_{0\mu}e_{j\mu}=0,
\label{23}
\ee
where indeces $0$ and $i,j=1,2,3$ are the tetrade ones, and 
the corresponding Christoffel symbols are
\be
\Gamma_{ij\alpha}=e_{i\mu}\frac{\pcd}{\pcd 
P_{\alpha}}e_{j\mu},\quad
\Gamma_{0j\alpha}=e_{0\mu}\frac{\pcd}{\pcd 
P_{\alpha}}e_{j\mu}.
\label{24}
\ee

It can be shown after some tedious algebra (see \cite{8,5} for the details)
that $P_{\mu}$ together with the variables
$$
Q_{\mu}=X_{\mu}+\frac12S_{ij}\Gamma_{ij\mu}
$$
\be
n_i=-e_{i\mu}\frac{x_{\mu}}{\sqrt{-x^2}}
\label{25}
\ee
$$
S_{ij}=e_{i\alpha}e_{j\beta}S_{\alpha\beta}
$$
commute in a familiar way,
\be
\begin{array}{l}
\dbp{P_{\mu}}{Q_{\nu}}=g_{\mu\nu},\quad\dbp{P_{\mu}}{P_{\nu}}=
\dbp{Q_{\mu}}{Q_{\nu}}=0,\\
{}\\
\dbp{n_i}{n_j}=0,\quad 
\dbp{S_{ik}}{S_{ab}}=-\delta_{ka}S_{ib}-\delta_{kb}S_{ai}-
\delta_{ia}S_{bk}-\delta_{ib}S_{ka},\\
{}\\
\dbp{S_{ik}}{n_j}=n_i\delta_{kj}-n_k\delta_{ij},\\
{}\\
\dbp{P_{\mu}}{n_i}=\dbp{P_{\mu}}{S_{ij}}=\dbp{Q_{\mu}}{n_i}=
\dbp{Q_{\mu}}{S_{ij}}=0,
\end{array}
\label{26}
\ee
so that variables (\ref{25}) are the physical ones of the spherical 
top \cite{9}. Namely, $Q_{\mu}$ is the four-dimensional analogue of the 
Newton--Wigner variable \cite{10}, whereas $n_i$ and $S_{jk}$ describe 
the internal angular motion.

Now we take remaining constraints $\f{3}(\beta)$  and $\f{6}(\beta)$ 
into account and define the final Dirac brackets as
\be
\{AB\}^*=\dbp{A}{B}-\sum_{m,n=3,6}\int_0^1d\beta_1\int_0^1d\beta_2
\dbp{A}{\f{n}(\beta_1)}\tilde{C}^{-1}_{nm}(\beta_1,\beta_2)
\dbp{\f{m}(\beta_2)}{B},
\label{27}
\ee
where 
$$
\tilde{C}_{mn}(\beta_1,\beta_2)=C_{mn}-\sum_{i,j=1,2,4,5}
\dbp{\f{n}(\beta_1)}{\f{i}}C^{-1}_{ij}\dbp{\f{j}}{\f{m}(\beta_2)}.
$$

Matrix $\tilde{C}_{mn}(\beta_1,\beta_2)$ is degenerate because of the presence
of the first class constraints (\ref{20}), so the integrals in (\ref{27})
are to be understood symbolically: for example, one can discretize the 
continious sets $\f{3}$ and $\f{6}$, replacing the integration over $\beta_1$
and $\beta_2$ by the finite summation, and exclude one pair $\f{3}(\beta_0)$
and $\f{6}(\beta_0)$ with an arbitrary $\beta_0$. We don't need to put this
procedure onto more rigorous grounds, because the final brackets for the
physical variables coincide with the preliminary ones. Indeed with the 
help of brackets (\ref{16}) one finds that 
\be
\tilde{C}_{33}(\beta_1,\beta_2)=0,\quad\tilde{C}_{66}(\beta_1,\beta_2)=0
\ee
and only terms containing $\tilde{C}^{-1}_{36}$ and $\tilde{C}^{-1}_{63}$
contribute to brackets (\ref{27}). 
This means that for the preliminary brackets not to be distorted it is 
enough to show
that $\dbp{A}{\f{3}}$ is zero for all $A$ belonging to the set of physical 
variables. As soon as $\{A\f{3}\}=0$ for 
any physical variable $A$, then
\be
\dbp{A}{\f{3}}=-\{A\f{2}\}C_{25}^{-1}\{\f{5}\f{3}\},
\ee
whereas $\{A\f{2}\}=0$ for all variables from set (\ref{25}). So finally
one has $\{AB\}^*=\dbp{A}{B}$ for all physical variables (\ref{25}).

Now, when the final Dirac brackets for the physical variables are established,
the redundant variables can be expressed in terms of physical ones by means of 
constraint surface equations $\f{a}=0$ with the result 
\be
\nu(\beta)=\frac{N}{\sqrt{1-(2\beta-1)^2}},\quad m=\frac{\pi N}{16},\quad
M=\frac{\pi N}{2},\quad k=\frac{\pi\sigma^2}{4N},
\label{29}
\ee
$$
x^2=-\frac{8L}{\pi\sigma},\quad p^2=-\frac{\pi\sigma L}{8},
$$
where $N$ is arbitrary, and $L^2=\frac12S_{ik}S_{ik}$. The presence
of an arbitrary constant in (\ref{29}) is the consequence of the first class
constraints (\ref{20}). Physically significant is the trajectory constraint
\be
H_0=-\frac{1}{2M}(P^2-2\pi\sigma L)\approx 0.
\label{30}
\ee 

To quantize the theory we are to find an operator realization of algebra 
(\ref{26}). For the centre-of-mass motion it is achieved with 
$\hat{P}_{\mu}=-i\frac{\partial }{\partial Q_{\mu}}$ in the coordinate 
representation, and the internal motion is described in terms of the angular 
momentum operator $\hat{L}_n=\frac12\varepsilon_{nik}\hat{S}_{ik}$ acting at
the components of the unit vector $n_k$ in the tetrade 3-space.

Trajectory constraint (\ref{30}) as the first class one leads to the equation
for the wave function,
\be
\left(\hat{P}^2-2\pi\sigma\sqrt{\hat{L}^2}\right)\Psi=0,
\ee 
with the spectrum
\be
P^2=2\pi\sigma\sqrt{L(L+1)}.
\label{32}
\ee

Alternatively, a gauge in the $\tau$-reparametrization group can be fixed,
{\it e.g.} by setting
\be
Q_0=\tau
\label{33}
\ee
(laboratory gauge). As the centre-of-mass is properly separated out, the 
quantization leads to the Schroedinger-type equation
\be
\sqrt{\vec{P}^2+2\pi\sigma\sqrt{L(L+1)}}\psi=E\psi
\ee 
with no ordering ambiguities. Other ways of gauge fixing (proper-time gauge,
light-cone gauge) can be used as well.

Nevertheless, it is not straightforward to fix the gauge at the level of the
Lagrangian. The standard method of quantization on some hypersurface does not 
work. Indeed, there is only one gauge group, and the gauge should be fixed by 
imposing only one extra constraint, like (\ref{33}). As there are two
single-particle coordinates $x_{1\mu}(\tau)$ and $x_{2\mu}(\tau)$  
at our disposal, two conditions are usually imposed in such type of problems
\cite{2}. For example, a popular choice is
\be
x_{10}(\tau)=x_{20}(\tau)=\tau.  
\label{35}
\ee
These two conditions are more than one gauge fixing constraint, and the 
resulting theory differs from the original one. In simple case (\ref{2}),
(\ref{3}) conditions (\ref{35}) satisfy the classical equations of motion
of the original theory, but with (\ref{35}) the motion is restricted to the 
rotations in the plane orthogonal to the three-dimensional vector $\vec{P}$,
and the quantization leads to the wrong Regge trajectory
\be
P^2=2\pi\sigma L
\ee
instead of (\ref{32}). Moreover, it is not clear {\it a priori} whether
conditions of type (\ref{35}) do not violate the equations of motion of 
the original theory in the case of straight-line string with massive ends 
(\ref{1}). To the contrary, the suggested formalism allows not only to 
establish unambiguously the Newton--Wigner variable $Q_{\mu}$ and the
corresponding internal variables, but also to fix the $\tau$-reparametrization
gauge in physically transparent and convenient way.
\smallskip

This work is supported by grants 96-02-19184a and 97-02-16404 of Russian
Fundamental Research Foundation and by INTAS 94-2851 and 93-0079ext.

\end{document}